# IS MANY LIKELIER THAN FEW? A CRITICAL ASSESSMENT OF THE SELF-INDICATION ASSUMPTION

Milan M. Ćirković

*Astronomical Observatory Belgrade*

*Volgina 7*

*11000 Belgrade*

*YUGOSLAVIA*

*e-mail:* `arioch@eunet.yu`

# IS MANY LIKELIER THAN FEW? A CRITICAL ASSESSMENT OF THE SELF-INDICATION ASSUMPTION

**Abstract.** We analyze the arguments allegedly supporting the so-called Self-Indication Assumption (SIA), as an attempt to reject counterintuitive consequences of the Doomsday Argument of Carter, Leslie, Gott and others. Several arguments purportedly supporting this assumption are demonstrated to be either flawed or, at best, inconclusive. Therefore, no compelling reason for accepting SIA exists so far, and it should be regarded as an *ad hoc* hypothesis with several rather strange and implausible physical and epistemological consequences. Accordingly, if one wishes to reject the controversial consequences of the Doomsday Argument, a route different from SIA has to be found.

## 1. Introduction: The Doomsday Argument and SIA

Probably the most vigorously discussed application of the anthropic reasoning during the last decade has been the Doomsday Argument (henceforth DA). DA was conceived by the astrophysicist Brandon Carter in early 1980s (unpublished), and it has been first exposed in print by John Leslie (1989) and in a *Nature* article by Richard Gott (1993). The most comprehensive discussion of the issues involved is Leslie's monograph *The End of The World* (Leslie 1996). The core idea can be expressed through the following urn ball experiment. Let two large urns are put in front of you, and you know that one of them contains ten balls and the other a million, but you are ignorant as to which is which. The balls in each urn are numbered 1, 2, 3, 4 ... etc. Now you take a ball at random from the left urn, and it shows the number 7. Clearly, this is a strong indication that that urn contains only ten balls. If originally the odds were fifty-fifty (identically-looking urns), an application of Bayes' theorem gives you the posterior probability that the left urn is the one with only ten balls as $P_{post}$ (n=10) = 0.99999. But now consider the case where instead of the urns you have two possible models of humanity, and instead of balls you have human individuals, ranked according to birth order. One model suggests that human race will soon go extinct (or at least that the number of individuals will be greatly reduced), and as a consequence, the total number of humans that will have ever existed is about 100 billion. The other model indicates that humanity will colonize other planets, spread through the Galaxy, and continue its existence for many future millennia; consequently, we can take that number of humans in this model is of the order of, say, $10^{18}$. As a matter of fact, you happen to find that your rank is about sixty billion. According to Carter and Leslie, we should reason in the same way as we did with the urn balls. That you should have a rank of sixty billion or so is much more likely if only 100 billion persons will ever have lived than if there will be $10^{18}$ persons. Therefore, by Bayes' theorem, you should update your beliefs about mankind's prospects and realize that an impending doomsday is much more probable than you have previously thought.

This is the original, Carter-Leslie version of DA. The version of Gott (1993) is somewhat different, since it does not deal with the number of observers, but with intervals of time characterizing any phenomena (including humanity's existence). Where Gott does consider the number of observers,[1] his argument is essentially temporal, depending on (obviously quite speculative) choice of particular population model for future humanity. It seems that a gradual consensus has been reached about inferiority of this version compared to Leslie-Carter's (see especially Caves 2000; Olum 2001), so we shall concentrate on the latter.

Its underlying idea is formalized by Bostrom (1999, 2000, 2002) as the Self-Sampling Assumption (henceforth SSA):

> **SSA**: One should reason as if one were a random sample from the set of all observers in one's reference class.

In effect, it tells us that there is no substantial qualitative difference between doing statistics with urn balls and doing it with intelligent observers (in this sense it is a "Copernican" assumption). SSA has several semingly paradoxical consequences, which are readily admitted by its supporters; in particular, see Bostrom (2001). Some of them, including alleged backward causation, as well as reality of future and/or causally disconnected observers, will be discussed in the next section of the present manuscript. Others are strong motivation for further work in this area. In particular, the **reference class problem** ("what counts as an observer?") has been plaguing the entire field of anthropic reasoning. A possible response to it is an improved version of SSA, proposed by Bostrom (2000) as "Strong SSA" (SSSA):

> **SSSA:** One should reason as if one's present observer-moment were a random sample from the set of all observer-moments in its reference class.

Although introducing important new elements in the discussion, in particular possibility of connection with cognitive sciences, SSSA does not change the essential conclusion of DA.

What are main arguments for accepting these sampling assumptions? As emphasized by Bostrom (2001, 2002, and references therein), there are both theoretical and practical reasons for accepting a statement equivalent or similar to SSA or SSSA. Theoretical reasons are manifested in plausibility of reasoning guided by the sampling assumptions in most conceivable situations where probabilistic reasoning about a set of intelligent observers is required (for a taste of specific examples, see *The Dungeon* or *The Incubator* thought experiments of Bostrom 2001). In other words, subjective credences assigned by using these assumptions maximize one's gain in any appropriate bettings. The practical side is even more interesting from our present point of view. In absence of any detailed understanding how intelligent observers emerge in the universe, we should not hurry to the conclusion that we are in any particular way "special". This (possibly qualified by further pieces of information) **typicality** is the essence of the sampling assumptions, and it has so far served quite well in many areas of application: notably cosmology, but also philosophy of physics, biology, quantum physics and even traffic planning!² This practical side of the issue has the additional benefit of emphasizing that these assumptions are not to be taken dogmatically, but will probably experience further qualifications and modifications as our knowledge on the physical, chemical, biological, etc. preconditions for observership grows (one such improvement has, in fact, already been suggested in Chapter 10 of Bostrom 2002).

However, an entire new strategy has been first suggested by Dieks (1992), and further developed by Kopf, Krtous and Page (1994), Bartha and Hitchcock (1999) and, in particular, Olum (2001). It is based not on refuting SSA (or SSSA), but on **adjoining** another general anthropic assumption that would compensate for the Bayesian probability shift in DA. Roughly speaking, the idea is that an observer is more likely to observe anything (i.e. to find herself alive) if there is a large collection of observers compared to the case of a small collection of observers. This assumption has, surprisingly, not been precisely defined in the papers proposing or defending it, but Bostrom (2000)—who criticized it—proposed a name Self-Indication Assumption (henceforth SIA), as well as the definition we may use for a start:

> **SIA:** Given the fact that you exist, you should (other things equal) favor hypotheses according to which many observers exist over hypotheses on which few observers exist.

Other locutions used in the literature (e.g. "one is more likely to find oneself in the long-lived race", Olum 2001) are equivalent to this. The fact that SIA **exactly** compensates for the DA-inducing probability shift has been demonstrated by Kopf et al. (1994), and it is certainly beyond doubt. However, the other merits and demerits of SIA have remained an open question; it has been criticised by Leslie (1996) and Bostrom (1999, 2000). As a recent interesting study by Olum (2001) indicates, the criticism has not been decisive, and SIA certainly remains—although unwarranted, as will be, hopefully, shown below—a strong competitor in the field of anthropic reasoning.

It should be immediately noticed that SIA is a *ceteris paribus* statement. Thus, rejecting SIA does not mean that we should prefer hypotheses with fewer observers or that we should *a priori* disfavor hypotheses with large number of observers if any other inclination to think otherwise exists. This inclination may be the embodied, for instance, by relative simplicity of competing theories; it may well be that simpler theory implies more observers (we shall return to this point). In other words, one must separate the legitimate *a priori* bias in favor of simpler theories, from the alleged *a priori*—but conditional upon one's existence—bias in favor specifically of hypotheses that imply more observers just because they imply more observers. The latter is subject of our critical examination in this study.

In a recent manuscript, Dennis Dieks (2001) has listed a number of problems associated with DA, some of which boil down to SIA, although Dieks himself professes strong reservations towards the latter assumption. In particular, Dieks suggests that the information about time one lives at necessarily changes the prior probabilities (assigned to various hypotheses about the future) in such way that the updated prior probabilities exactly cancel with the DA probability shift. Now, this may be literally valid against Gott's version of DA, in which the **duration** of phenomena is crucial. However, in the case of Carter-Leslie version, in order to achieve the same result, one has to conditionalize upon the number of observers existing in various investigated cases (which is what Dieks actually does)—which is just a particular operationalization of SIA.

All this activity should not convey the impression that the only conceivable reason for taking SIA seriously is the answer to DA. This would be an oversimplification. *Prima facie*, SIA stands alone as an assumption in anthropic reasoning whose merits and demerits should be investigated. It should be kept in mind that the way of thinking embodied by SIA would be attractive for scientists and philosophers even in the absence of the DA challenge. It may have important consequences for currently hotly debated issues in quantum cosmology (Olum 2001). Therefore, a detailed critical examination is certainly a legitimate endeavor.

## 2. Arguments derived from problems in sampling assumptions

Let us first consider some of the arguments listed against using the alternative assumption (either SSA or SSSA) **as sufficient** in the anthropic reasoning. It is legitimate to invoke such arguments in support of SIA as long as we are convinced that alternative assumptions have no recourses left. As we shall see, this conclusion is hardly warranted.

**2.1. Backward causation implied by SSA**

Olum (2001) cites examples of unreasonable predictive powers and alleged backward causation implied by SSA and related assumptions as an argument for complementing this mode of reasoning by SIA. Detailed and colorful exposition of such cases is given by Bostrom (2000, 2001), although the latter author apparently believes that these difficulties are surmountable. Let us here cite one of these examples:

> It is the year 2100 A.D. and technological advances have enabled the formation of an all-powerful and extremely stable world government, $UN^{++}$. Any decision about human action taken by the $UN^{++}$ will certainly be implemented. However, the world government does not have complete control over natural phenomena. In particular, there are signs that a series of *n* violent gamma ray bursts is about to take place at uncomfortably close quarters in the near future, threatening to damage (but not completely destroy) human settlements. For each hypothetical gamma ray burst in this series, astronomical observations give a 90% chance of it coming about. However, $UN^{++}$ raises to the occasion and passes the following resolution: It will create a list of hypothetical gamma ray bursts, and for each entry on this list it decides that if the burst happens, it will build more space colonies so as to increase the total number of humans that will ever have lived by a factor of *m*. By arguments analogous to those in the earlier thought experiments, UN$^{++}$ can then be confident that the gamma ray bursts will not happen, provided *m* is sufficiently great compared to *n*.

At first glance, the same application of SSA as in DA suggests that γ-ray bursts will not have occured, because of their **future** consequences (vast increase in the number of observers). Along the line of thought suggested by Bostrom, we shall try to show that the appeal to unreasonable predictive powers and/or backward causation as arguments against DA and SSA is based on at least inconclusive premises.

  First of all, long philosophical tradition suggests that there is no reason to reject backward causation *a priori*, in particular if one believes the so-called B-theories of time to be correct.[3] In fact, as demonstrated in a recent brilliant monograph of Price (1996), backward causation may even be necessary for the completely coherent explanation of the conventional arrows of time (thermodynamic, radiative, etc.). There is no compelling reason for rejection of backward causation as long as one is able to escape the so-called "bilking" paradox, representing the root of all causal problems with time inversion or time travel to the past familiar from many films and science fiction stories. This sort of paradox arises when one claims that it is possible to bring about an earlier event A through a later event B. To see the paradoxical consequences of such correlation between future and past, it is enough to introduce an additional event C which occurs as a consequence of A and which **prevents** B from occuring. In already conventional terms, the correlation between A and B is bilked by C (Flew 1954; Mellor 1981). This problem lies in the background of all stories about an assassin travelling to the past and killing one of his ancestors, which prevents him from existing and travelling to the past, etc. What is important to notice is that the bilking paradox assumes a particular **capacity** of the bilking agency (e.g. the possibility of time-traveller to go to his own past and acts there). This is also transparent in the famous example of Newcomb's game "paradox": seemingly paradoxical results follow only from a very specific and non-standard **agency** of the being which creates the game setup (e.g. Schmidt 1998).

  It seems quite clear that there is no danger of bilking paradox as far as SSA-related issues are concerned. The conclusions drawn from the possible existence of large or small number of future observers do not remove a factor necessary for their existence, i.e. there is no relevant bilking agency. It is impossible to set up the event C which will correlate with ¬B in the DA case. Therefore, although one may classify this as a case of backward causation, there is no appropriate paradoxical consequences.[4] This mode of defense of what formally represents backward causation is rather well-known in the philosophical literature, and has been recently colorfully exposed by Brown (1992) in the context of the time-travel problem, relying on several arguments previously put forward by Dummett (1954). Similar argument has been used by theologians and philosophers of religion to show that the human freedom of will does not cause paradoxes when iuxtaposed with the Divine perfect foreknowledge (cf. Craig 1988).

The second relevant issue is that there is a sort of semantic confusion here. When Olum (2001) states that SIA "agrees with our intuition that the chance of an event (e.g., the earth being hit by an asteroid) should not depend on the event's consequences (e.g., humanity being wiped out)" he seemingly ignores what he has written only a few pages earlier, that "there are different kinds of probability". There he explicitly distinguishes between objective physical probabilities, and the subjective ones following from our ignorance. In the asteroid case, we are obviously dealing with the latter, since it is far from clear that we understand all factors playing a role in the occurences of catastrophic impacts in earth's history. Besides, any appeal to our intuition is unfortunate when temporal vs. causal relationships are concerned; as correctly warned by Price (1996, p. 81), "in this area, more than most, our intuitions are a poor guide to the explanatory priority".

## 2.2. Reality of future and possible observers

Critics of the "sampling" assumptions' (SSA, SSSA) sufficiency and their doomslike consequences like to point out that future and/or merely possible observers are not to be treated equally (or even not at all) as the past and present observers. This sort of argumentation seems removed from the physical point of view. It is well-known that fundamental equations of physics (with minuscule exception of CP violations in some rare weak decays) know nothing about the direction of time. This was one of the motivations for introducing B-theories of time mentioned above, in which there is no temporal becoming, and therefore no objective ontological distinction between future and past observers. The distinction we tacitly imply in everyday's discourse has to do with our perception of the world and with our language, not with the physical structure of the universe (Price 1996). Time neither flows, nor do things come into existence except in the trivial sense that we are conscious of them at one moment after not having being conscious of them at any earlier moment. Thus, the objection to SSA and related assumptions based on different objective status of future observers dissolves on B-theories of time. Now, one may argue that B-theories are inadequate for the task, but it is important to understand the price which comes with such an attitude. Much of the practical physical work is based on the assumption that world-lines of particles within each body constitute a total and unchangeable 4-dimensional shape of everything that happens with that body. To reject this notion means to reject a large part of modern physics, including the special theory of relativity, as fundamentally incomplete. This is the reason why even opponents of the B-theory admit that it may be relevant in respect to the physical time (which we are interested in the present context), while asking only that the full metaphysical treatment include the temporal becoming.[5] The same reasoning applies to treatment of possible observers according to SSA, since in any particular situation considered there are no possible observers in (our perspectival) past, only in (perspectival) future.

## 2.3. Causally disconnected influences

There are two sorts of influences allegedly coming from causally disconnected regions: probability shifts due to observers which are not currently causally connected with us at present (technically speaking, those which are beyond the *particle horizon*), and those due to observers which will remain causally disconnected from us **at all times** (those located beyond the *event horizon*).[6] Our epistemic capacities are different in these two cases, and they should not be confused or mixed together as is usually done. For the probabilistic arguments as discussed in the introductory section, the former do not present a particular difficulty, exactly for the reasons discussed in previous two subsections. The latter may pose a difficulty, since the observers beyond the event horizon will stay there forever, and for any practical purposes they may as well live in a different, topologically disconnected universe. Their including in the total tally of observers in our universe is at least unclear at present (see the discussion in Sec. 4 below). This is particularly troublesome if one takes seriously the class of cosmological theories which explicitly violates the cosmological principle (homogeneity and isotropy) on very large—"superhorizon" in cosmological parlance—scales, as many modern inflationary models do. Consequences of this are serious enough: we may in this manner gain additional reason to include the values of cosmological parameters of our universe in any realistic universal assumption dealing with observership. Namely, the event horizon will arise in our universe mainly in the case (considered realistic since 1998) of its possessing a positive cosmological constant (Ćirković and Bostrom 2000). Thus, the status of this complaint against sampling assumptions based on causal disconnectedness is inconclusive, to say at least.

Here the table may in fact be turned and this restriction shown to be unfavorable to SIA, since if we accept only observers within the event horizon, this means that we automatically favor cosmological theories with smaller number of observers if everything else stays the same (horizon by definition encloses smaller spatial volume from the volume of the universe). Thus, the true prediction from SIA would be the absence of event horizons in our universe, which was widely accepted until a couple of years ago, but is now considered disproved on observational grounds.

## 3. Arguments against SIA

People who, like Leslie and Bostrom openly adopt different sets of methodological assumptions in the anthropic reasoning, have considered several arguments allegedly refuting SIA. Hereby we would like to add several new ones to these discussions. While we consider several possible arguments against SIA in this Section, we postpone the so-called "presumptuous philosopher" thought experiment until the Sec. 4. The latter thought experiment is so rich in its importance and consequences that it deserves a separate discussion. For other arguments see Chapters 5 and 6 in Leslie (1996) and Chapter 7 in Bostrom (2002).

### 3.1. SIA and other planetary systems

The presence of terrestrial planets elsewhere in the universe is, of course, of the paramount importance for any anthropic discussion. (It is so in particular since Bostrom (2000, Ch. 9) has shown that the "freak observers", i.e. those intelligent beings not evolved similar to ourselves, but resulting from accidental environmental fluctuations, cannot significantly influence our judgement on the likelihood of cosmological theories.) Leslie (1996) discusses Marochnik's (1983) theory about emergence of planetary systems only in the vicinity of the galactic corotation distance, and Olum (2001) critically examines the same example. However, Olum's discourse is at least confusing on the issue. While admitting that after we find ourselves close to the corotation distance of the Milky Way, we should look more favorably on that theory, he writes:

> If you feel that in advance of observation, the theory was wildly unlikely, then you can feel even though the evidence has made it much more likely, that it is still not an especially well-supported theory. The reason to feel that it is a priori unlikely, of course, is that it leads to a very small number of observers, as opposed to the theory in which planets are common everywhere. (You could also feel that this theory is unlikely, in advance of observation, because of the possibility that it will immediately be ruled out by finding that we are in a place where no planets should have been.)

First of all, the "explanation" why we should consider Marochnik's theory highly unlikely in the first place is just begging the question. Since this is only a reformulation of SIA, it cannot give the latter any additional weight. Secondly, the qualification in parentheses is at least a strange and highly non-standard criterion for judging a physical theory. By the same token, one might have stated in XVI century that Copernican theory of the solar system was unlikely, since a simple observation that Earth is at rest (via, say, Foucault's pendulum which could have easily been invented at the time) would decisively disprove it. Such counterfactuals are always true, but also not very informative! On the other hand, the criterion seems exactly contrary to the classical Popperian view of falsifiability as the main criterion for evaluating scientific theories. The theory which is easily falsifiable (as the Marochnik's one) is more scientific than a theory (like the one that there are many inhabited planets in galaxies other from ours, used by Olum), which is very, very hard to disprove. Historically speaking, the steady state theory did an immense service to cosmology exactly because it was much easier to disprove than any of the rival relativistic cosmological models (Kragh 1996).

There is some independent evidence based on investigation of the recently discovered extrasolar planets, that stars with planets (and therefore sites of observers' emergence) are indeed anomalous, as far as chemical evolution is concerned (Gonzalez 1999). This, if confirmed by subsequent observations, may have several important ramifications for the anthropic reasoning. In particular, it may help solve the long-standing problem of not perceiving cosmic "miracles", or any other manifestations of advanced extraterrestrials, known as the Fermi's "paradox". We shall return to this issue later on. For the moment, let us notice that on SIA, the hypothesis that stars with planets (and in particular the subset of stars with inhabitable planets) are in any way anomalous is *a priori* less favored than the hypothesis that all or almost all stars have planets. In the situation in which we are at present, knowing just minuscule bits about the physical requirements for the emergence of life and intelligence, empirical examples on which SIA and similar assumptions may be tested are scarce indeed. However, the example of the study of Gonzalez (1999) shows that SIA may be misleading, since it would suggest that we should—in the daily research business—concentrate on the subclass of planetary system formation theories assuming uniform formation of planetary systems. In this manner—if the finding of Gonzalez is confirmed by further observations—SIA would (mis)lead us to, at least, wasting a deal of theoretical effort through ignoring/downplaying an important aspect of the problem. It is not too far-fetched to argue, as we shall do below, that it is misleading in other areas too.

Similar things may happen in cosmology. Consider, for instance, another highly controversial cosmological theory, Tipler's Omega-point theory (Tipler 1986, 1994; Barrow and Tipler 1986). In it, the cosmological parameters are strictly constrained by the final boundary conditions of the closed universe (this example has the virtue of being spatially finite, and therefore completely tractable in view of additional anthropic assumptions). However, SIA is inapplicable even in the context of this theory (which should be more favorable to SIA than most of other cosmological theories, since it also suggests existence of a large numbers of observers overall, here motivated by teleological reasons). The value of the cosmological density fraction[7] $\Omega$ determines the time remaining until the "Big Crunch" and, therefore, the maximal spatial size of the universe. We would naively expect $\Omega$ to be very, very close to unity (while, of course, still larger than unity) in order to have a very big universe; but this fails, because final boundary condition precludes it. Also, if one accepts rescaling of the time scale more in accordance with the dynamics of physical processes, as suggested by Tipler, and investigated earlier by Misner (1969), than it is bizarre to have merging of larger and larger aggregates of observers in the infinite subjective time close to the final singularity, resulting in effectively fewer and fewer observers (while our new—rescaled—time axis is infinite).

## 3.2. SIA and teleology

A general problem with any approach affirming SIA is that it is to some degree necessarily reductionist. For instance, one can argue that there is no causal connection between propositions on the predicted number of planets and on existence of intelligent observers in any number of competing theories, unless one introduces an additional assumption which claims that once you have physical conditions within some prescribed range, the emergence of life and ultimately intelligence is inevitable. This is a rather strong assumption, at least while we know so little on the actual physical, chemical and biological preconditions for the existence of intelligent life. It is important to note that this is not reductionism *per se* which is dubitable here, but its **necessity** in the framework of such an approach. In general, such form of necessity appears mainly in the teleological context, and indeed, the entire set of arguments usually claimed for SIA has an air of teleology about it. If there are additional requirements for emergence of consciousness (and, contingently, observership) on the quantum level, as many modern physicists from Wigner to Wheeler to Penrose thought, than those would effectively modify all the relevant priors, and thus invalidate the "simplicity" of SIA. For instance, if Wheeler is correct in stating that observers are necessary to bring the universe into being (cf. Barrow and Tipler 1986) **and** we for some reason ascribe positive value to existence of larger and larger chunks of the universe, then it may be understood why the large number of such observers is desirable. But even then it is not the existence of observers *per se* which is important, but their specific capacity (of reducing the universal wavefunction), which invokes the complicated underlying and highly speculative physical mechanism.

### 3.3. SIA and Quantum Mechanics

In the Appendix to his monograph on DA, John Leslie has argued that DA refutes Everett's "many world" (or "no-collapse") interpretation of quantum mechanics (Leslie 1996, pp. 264-266). We do not wish to comment upon Leslie's argumentation here, just noting that it is based upon a particularly strong and literal version of SSSA, counting *all* possible observer-moments as admissible. One of the virtues of Bostrom's (2002) approach to the *reduction* of SSSA is exactly that it enables "filtering" of the undistinguishable observer-moments in the final tally. Therefore, Bostrom's version turns out indifferent towards the choice of the most adequate quantum-mechanical version (which agrees with our intuitions on the subject).

However, we would like to look at the issue of the "no-collapse" quantum mechanics from the point of view of SIA. First of all, such an interpretation is preferable over the rival Copenhagen or dynamical reduction or any other interpretation *ceteris paribus*.[8] SIA thus not only supports Everett's theory,[9] but will prefer particularly exotic and bizarre branches of the universal wavefunction in which the number of observers is much larger due to the extravagantly early structure formation, more convenient power spectrum, etc. This is related to another old debate in the anthropic reasoning. As noticed by several critics of the anthropic principle (e.g. Gould 1985), universe still *could* be much more hospitable to life and intelligence than we perceive it to be. It is reasonable to assume that in Everett's multiverse there are many such branches, which would be preferred by SIA. (In other words, the "no-collapse" theory provides a physical "vehicle" for substantial improvement of the observed universe in this—rather simplistically understood—sense of hospitability.) Therefore, the fact that we do see physical space which could, in principle (and in agreement with other physical constraints, such as energy conservation), be inhabited by large intelligent populations is once more the argument against SIA, this time based on the selection of particular branches of the universal wavefunction. In addition, since the most fatal illnesses (as well as the old age itself) are presumably reducible to the quantum-level phenomena, the reasoning behind so-called "quantum suicide" thought experiments (Squires 1986; Price 1996; Tegmark 1998) may apply to them, in which cases SIA would suggest not that we should find ourselves near death (*vide* Leslie), but that we should find ourselves—and other fellow humans—immortal as well!

Of course, a simple solution would be to claim that the "no-collapse" theory fails on other counts (for instance the definition of probabilities), so that any application of anthropic assumptions, be it SIA or SSA or any other, can not be legitimate. It does not seem an easy task, since the "no-collapse" view after more than four decades of its life shows all signs of vitality and even increasing popularity (cf. a rough poll in Tegmark 1998), and it seems especially appropriate in the quantum cosmological context (Barrow and Tipler 1986). If one needs to choose between discarding SIA and discarding the "no-collapse" theory, the former option looks significantly more attractive.

### 3.4. Davies-Tipler argument, entropy and SIA

SIA leads to fallacious predictions when it is applied to relatively small spatial or relatively large temporal scales. Isn't it obvious that it is more plausible that a typical planetary system is inhabited by $10^9$ than $10^{99}$ observers? Let us investigate in detail why should we think that sometimes small number of observers is preferable to a large number.

First of all, observers are here understood as physical entities, subject to physical laws, and in particular to the Second Law of thermodynamics. As shown by Szilard, Brillouin and others (e.g. Brillouin 1962), there are definite thermodynamical limitations on functioning on any information-processing devices, **including** intelligent observers. In particular, the maximal amount of processed information is linearly proportional to the energy invested in the processing itself in the ideal case (and in the real case, we expect consumption to be larger still). Astrophysically speaking, this means that there is (in principle calculable) an upper limit on the spatiotemporal density of processed information depending on availability of various astrophysical energy sources, like nuclear fusion in the stellar interiors, or gravitational collapse of massive stars or Penrose process of energy extraction from rotating black holes. In other words, the entropy production rate in a fixed comoving volume is necessarily limited, although as suggested by Dyson in his pioneering study in physical eschatology (Dyson 1979), an advanced civilization may wish to intentionally slow it down in the extreme to ensure its longevity (and in a special case consider by Dyson even immortality). This immediately gives an upper limit to a comoving density of observers in the universe.[10] True limits are, of course, much more stringent, since the entropy production is always larger than the optimal value.

This does not address only a rather trivial assertion that SIA *prima facie* supports senseless hypotheses like the one of $10^{99}$ observers per planetary system. This also answers another issue raised in the discussions of DA and related hypotheses, namely that we should limit ourselves to the spatial region which is observable. In that region, the probability distribution function of observers *p(N)* is non-zero only on a finite interval (0, $N_{max}$), where $N_{max}$ is the theoretical maximal number of observers. Moreover, this function has to behave in such a way as to take very small values where N is slightly smaller than $N_{max}$ (extraordinary optimization required). Insisting on SIA means that *p(N)* more or less monotonically increases with N; this, in turn, implies that the efficiency of energy extraction for purposes of information processing increases with the number of observers present in a causally connected region of spacetime. This is an anthropic constraint closely related to Fermi's "paradox", which was discussed, among others, by Paul Davies and Frank Tipler—in particular in connection with the cosmological models with infinite past, but this can be also applied (in a weaker form) to finite-past universes like ours. The fact that we have not yet found any physical reason for non-existence of supercivilizations billions of years older than ours in the Galaxy, and we still do not perceive even the slightest traces of their computational processes and unavoidable entropy increase (not to speak of their physical absence from Earth and the Solar system), indicates subtler reasons for preferring ¬SIA to SIA.

Actually, this is just the second horn of the dilemma. If we wish to avoid problems with infinity, we need (as Bostrom 2000 correctly suggests in his footnote 83) to analyze the relevant densities. However, too high densities are more implausible than low densities, as we have argued above, and SIA is *at least* of limited application. But the total number of observers equals the density of observers times the total volume in a particular cosmological model. If we insist on SIA in finite universes, this can only mean that *ceteris paribus* we assign larger and larger probabilities to spatially larger and larger universes, which means that (in the context of standard relativistic cosmological models) we favor smaller and smaller $\Omega$, while it must be still larger than unity. In accordance with the reasoning above, there has to be a reasonably small positive value of $\varepsilon$ such that $\Omega = 1+\varepsilon$ corresponds to the maximum of the probability function. This strange method of learning cosmological facts without doing observations or simulations leads us directly to our next topic.

## 4. The Return of the Presumptuous Philosopher

Let us again consider the question at the bottom of most of the anthropic thinking: what does qualitatively distinguish intelligent observers (or their observer-moments) from urn balls, or any other simple probabilistic system? The exact answer to this question depends, of course, on the physical, chemical, biological or cognitive preconditions for observership which we know very little about. In the meantime, while acknowledging our ignorance, we operate with observers as with balls, **with some additional assumptions**. Some of them are of general *a priori* nature (SIA, SSA, etc.), while others may be physically or sociologically based. As we have seen, a part of this additional input is the requirement that the density of observers is **smaller** than some critical value, following from the Davies-Tipler argument. Thus, SIA certainly has a limited validity—although it gives higher credence to theories (whatever they might be) predicting $10^{100}$ observers per planetary system than those predicting only a couple of billion, it is quite obvious which of the two is more likely to be correct. With this in mind, we may approach the central SIA-related thought experiment.

The best litmus test for various related concepts is the beautiful thought experiment invented by Bostrom (2000) under the title of "presumptuous philosopher". The set of ideas embodied in this gedanken and its possible modifications is truly remarkable and encompasses almost all aspects of the anthropic reasoning. Let us quote the original version before discussing some modifications (Bostrom 2000):

> **PP1:** "It is the year 2100 and physicists have narrowed down the search for a theory of everything to only two remaining plausible candidate theories, T1 and T2 (using considerations from super-duper symmetry). According to T1 the world is very, very big but finite, and there are a total of a trillion trillion observers in the cosmos. According to T2, the world is very, very, *very* big but finite, and there are a trillion trillion trillion observers. The super-duper symmetry considerations seem to be roughly indifferent between these two theories. The physicists are planning on carrying out a simple experiment that will falsify one of the theories. Enter the presumptuous philosopher: "Hey guys, it is completely unnecessary for you to do the experiment, because I can already show to you that T2 is about a trillion times more likely to be true than T1 (whereupon the philosopher runs the God's Coin Toss thought experiment and explains Model 2 [which is SIA])!"

Bostrom finds such a posture deeply unfounded, even senseless (hence the qualification of the philosopher as "presumptuous"!), thus showing the unacceptable claims directly following from SIA. However, there is more here than catches the eye, and therefore we dub the setup in this original version for the sake of brevity PP1.

There are several crucial issues in this type of anthropic reasoning which are highlighted by this thought experiment. First of all, there is a condition of finiteness of the universe, to which we shall return later. Olum avoids to discuss this problem, since one of the more bizarre consequences of SIA is that, by extension, any universe containing an **infinite** number of observers is immediately infinitely more likely than any theory predicting only finite number of observers. In other words, finite theories are **already** discarded, since one may argue that there is at least one theory predicting infinite size of the universe and satisfactorily explaining all other cosmological data (the *ceteris paribus* clause being thus satisfied). Of course, one may put the definitional proviso, in the manner of Bostrom (2000) that infinite case is *a priori* excluded. However, one can hardly escape the impression that the limiting process of bigger and bigger universe will correctly approach the infinite case (at least of the smallest infinite cardinality). Therefore, the finiteness assumption requires some explanation.

But the deepest set of issues stemming from the presumptuous philosopher thought experiment is motivated by the following question: **what exactly are the theories T1 and T2 considered here?** This is not at all obvious. Bostrom in the original exposition (our version PP1) explicates that these are candidates for "The Theory of Everything", a label theoretical physicists often attach to the envisaged unified field theory incorporating all our knowledge on the fundamental constituents and forces of the universe (e.g. Weinberg 1993). Such a theory would be able *in principle* to derive any aspect of the observable reality from its theoretical structure (including, presumably, a—hopefully small—number of free parameters), in the same manner in which Laplace and other classical physicists believed the Newtonian theory to be able to predict and retrodict *in principle* all events in the entire history of the universe (cf. Feinberg, Lavine and Albert 1992). Note that after we accept a very weak reductionist assumption that there is a finite chance for life and intelligence to spontaneously evolve in places with favorable conditions, the only prediction we actually extract from T1 and T2 is the *size* of the universe. In other words, we calculate a number of cosmological parameters like the total matter density $\Omega$, the cosmological constant (presumably zero or negative) $\Lambda$, the Hubble parameter $H_0$ and a possible number of other parameters describing the difference between the real case and the simple Friedmann model. Now, this is a surprisingly small amount of information from entire fabulous wealth the true TOE could offer; in other words, T1 and T2 *need not* be the true TOEs, as long as they are the correct—in the sense of being self-consistent and consistent with other relevant physical theories— *cosmological* theories, or some wide (but not as wide as the true TOE) theories incorporating the realistic cosmological model. The true TOE would contain an enormous amount of information beside the simple information on the spatiotemporal size of the universe. In fact, one could convincingly argue that the true TOE would be able to predict *in principle* the details of emergence of any one single community of observers and even any one individual observer! This is not any more surprising or far-fetched than the classical Laplacian determinism.

Therefore, there seems to be a strange redundancy on PP1. We have two candidates for the final theory, both of which could, at least in principle, even if not in practice due to computational limitations, predict the exact number of observers in the universe. Since this is a thought experiment, one should not worry too much about these practical computational limitations; it is enough to suppose that such information is hidden in the equations of both candidate TOEs. Furthermore, it is plausible to assume that the comparison between the two theories would be a much simpler task than actually calculating the predicted number of observers in each theory. However, in PP1 we choose not to pursue this prediction, but instead rely in the assumption of uniformity of physical, chemical and biological processes leading to emergence of life and intelligence under similar conditions in various parts of the universe, and then compare these highly uncertain, coarse-grained averages. Such procedure is arguably much more feasible in computational practice, although it may be noted that for the purposes of the thought experiment this does not make big difference. Instead, the procedure employed in PP1 introduces not only idea of large-scale uniformity, but also a subtle assumption much discussed since Boltzmann, that any consistent universe must be at least as large as we perceive through the best contemporary telescopes, and that seems much larger than it is necessary for emergence of life and intelligence on Earth.[11]

Motivated by these expectations from the true TOE, let us consider slightly different version of the same gedanken experiment, which we shall dub PP2:

> **PP2:** "It is the year 2100 and physicists have narrowed down the search for a theory of everything to only two remaining plausible candidate theories, T1 and T2 (using considerations from super-duper symmetry). According to T1 the world is very big but finite, and contains about trillion trillion observers. According to T2, there are a trillion trillion trillion observers in the universe. The super-duper symmetry considerations seem to be roughly indifferent between these two theories. The physicists are planning on carrying out a simple experiment that will falsify one of the theories. Enter the presumptuous philosopher: "Hey guys, it is completely unnecessary for you to do the experiment, because I can already show to you that T2 is about a trillion times more likely to be true than T1!"

What are the effects of leaving the size of the universe out of the picture? The most important gain is that we do not need to worry about the uniformity of processes leading to the development of life and intelligence on very large scales. There is a host of contemporary cosmological models in which the universe is wildly inhomogeneous at scales larger than our particle horizon (especially inflationary models). PP1 is clearly inadequate to cope with the situation in such models, since no theory predicting just the size of the universe can tell you anything about the number of observers there if you do not accept an additional assumption describing uniformity of physical conditions. For instance, let us consider the theory (T2') in which the universe is closed, but is $10^3$ times larger (in linear scale) than our particle horizon and galaxies exist only in $10^{-6}$ of the total volume, while the rest is filled with black holes. How does this theory fare against the alternative (T1') that the universe is 10 times larger than our present horizon and completely uniform? If the size of inhabitable reason is the only generator of intelligent observers (i.e. there are none of Bostrom's "freak" observers), the two theories are completely equal. On the other hand, the presumptuous philosopher in PP1 could still argue for T2' with $10^6$:1 odds.

The problem which remains in PP2, is that it still does not make any difference between observers in observable and those in unobservable regions of the universe. Also, the problem of necessary finite universe remains with us. Let us, finally, consider the third version of the same experiment:

> **PP3:** "It is the year 2100 and physicists have narrowed down the search for a theory of everything to only two remaining plausible candidate theories, T1 and T2 (using considerations from super-duper symmetry). According to T1 the density of intelligent observers is about trillion trillion observers per cubic Megaparsec at present day. According to T2, there are on the average a trillion trillion trillion observers per cubic Megaparsec at present day. The super-duper symmetry considerations seem to be roughly indifferent between these two theories. The physicists are planning on carrying out a simple experiment that will falsify one of the theories. Enter the presumptuous philosopher: "Hey guys, it is completely unnecessary for you to do the experiment, because I can already show to you that T2 is about a trillion times more likely to be true than T1!"

By this reformulation, we achieve a substantial gain in clarity, since it is now legitimate not to worry about causally disconnected regions at all. If we abide by the condition that rival theories are true TOEs, it is completely legitimate to compare their predictions on the number of observers within our current particle horizon, i.e. predicted observer densities. Now, it is obvious that the reasoning of our philosopher must be wrong if the numbers are allowed to vary in a very large range. For instance, T2 seems disproved even now, if the number trillion trillion trillion (i.e. $10^{36}$) applies to our present horizon. One should keep in mind that the total number of baryons within horizon is only close to $10^{80}$, and that number of galaxies is of the order of $10^{11}$. It is hard to imagine that there can be $10^{25}$ observers per galaxy (and much more in the Milky Way, since it is an exceptionally large galaxy in the set of all observable galaxies); as far as our Galaxy is concerned, this estimate implies about $10^{15}$ or more observers per inhabitable planetary system, which seems wildly implausible, even before Fermi's "paradox" is taken into account.

Another gain is that we may drop the qualification of universe as finite. Now we may as well discuss infinite universes, with infinite total number of observers, while the density is everywhere finite. The true TOE is expected to predict **both** size of the universe **and** the spatiotemporal density of observers. These two predictions should **not** be regarded as redundant, but rather complementary. PP3 is advantageous still more since a critic could argue against implausible conclusion of PP1 and PP2 that the conjecture (supported by all current observational data![12]) that our universe actually is infinite automatically invalidates the entire chain of reasoning. PP3 enables us even to compare finite (but very large) and infinite cosmological models on even grounds. Hence we easily avoid the problem present in PP1 that SIA suggests that an infinite universe is infinitely times more probable than **any** finite universe (i.e. that the universe cannot be closed).

As argued above, in situations where constant comoving volume is considered, it seems more appropriate to argue for smaller number of observers than for the larger one. In the manner of Hempel's "raven paradox" (a white sheep observed **does** lend some credence to the hypothesis that all ravens are black!), your existence (and entropy consumption thereof) in a finite spatial volume suggest anti-SIA rather than SIA.

It should be emphasized that it is hardly possible to treat infinite universes without introducing the horizon limitation. If humankind is not entirely miraculous, there is a finite probability density for a technological civilization to arise at each point of spacetime. By virtue of the cosmological principle, all these probability densities, when averaged over a large enough volume, must be the same. In an infinite universe, this immediately means that the number of technological civilizations is infinite. However, the infinite number of civilizations entails multiple copies of every civilization, including multiple copies of ourselves as observers (Ellis and Brundrit 1979). Thus, **all relevant** reference classes are infinite, and all auxiliary anthropic assumptions become vacuous: in the same sense as it seems fallacious to derive probability shifts from any finite rank in an infinite set of observers (SSA), it seems that there is no predictive power in SIA if the set of observers is (regardless of the actual theory used) is infinite. How do we compare theories if we know that there is an actual infinity of observers in any case?

(And even if we do not trust the cosmological principle, it is enough to accept much weaker assumption that variations between spatial regions are not too large to prevent intelligence from arising in all but infinitesimal fraction of all spacetime. Since intelligent communities/civilizations are discrete entities, the relevant densities cannot asymptotically approach zero, and therefore any sum over infinitely many regions must diverge.)

## 5. Some Indirect Arguments

In his study, Olum (2001) points out that SIA also suggests some extremely implausible epistemological consequences. His example of such pathological behavior is worth citing in full, since it indicates one of the weakest points in the defense of this assumption:

> For example, suppose I have a crazy theory that each planet actually has $10^{10^{100}}$ copies of itself on "other planes". Suppose that I (as cranks often do) believe this theory in spite of the fact that every reputable scientist thinks it is garbage. I could argue that my theory is very likely to be correct, because the chance that every reputable scientist is independently wrong is clearly more than 1 in $10^{10^{100}}$. To avoid this conclusion, one must say that the a priori chance that my theory was right was less than 1 in $10^{10^{100}}$. It seems hard to have such fantastic confidence that a theory is wrong, but if we don't allow that we will be prey to the argument above.

It should be added here that this degree of confidence is trillions of trillions times higher than the degree of confidence we may have about wrongness of ideas postulating ghosts, aether, flogiston, extraterrestrial origin of Mayas and Egyptians (or of today's philosophers!), geocentrism, etc. Olum's subsequent attempt to show that the paradoxical consequences of SIA are encountered in everyday life without anthropic background is flawed (apart from obviously being *quid pro quo*). Since Pascal, we know that it is not the odds *per se* that is interesting in any actual betting, but the payoff also. Olum writes:

> However, similar scenarios exist without any dependence on the number of observers. For example, suppose a stranger comes up to you with the claim that if

> you give him a dollar today he will give you $10 tomorrow. Presumably you won't give him the dollar, which shows (if you are maximizing your expectation) that you think the chance he will come through as he says is less than 10%. On the other hand, it would be strange to claim that the chance is less than one in a million, since sometimes people making statements like this are honest. At this level you might even consider the possibility that your whole understanding of the world has one chance in a million to be very wrong, and so you can't trust your expectation that you won't be paid to this level. Nevertheless, if the payoff is raised to $10 million, you still won't give the dollar, which shows that now you think the chance for a payback is in fact less than 1 in 10 million.

The last sentence is strange to qualify. The present author would be happy to give a dollar in described circumstances, provided that *ceteris paribus* apply, i.e. that the stranger does not give some other circumstantial reason for doubting his intentions (like wearing a mask, or being obviously insincere on some other issue). As correctly pointed out by Olum (personal communication), this example is loaded with social and economical context and may not be the correct analogy. It would be better to consider a complicated hazard game, or a slot machine with no epistemic or ethical capacities, only success/failure options. Then the social issues (like material inequality of people or all-too-human tendency to cheat) do not play any role, and our hypothesis may be stated in the following manner: *One should be more ready to invest $1 into a game machine (with unknown internal structure) if its specifications promise $1,000,000 than in the case they promise only $10.* Thus, we escape the implication of Olum that there must be some deep underlying probability distribution which is not apparent, but may be constrained through analyzing our (intuitive!) chances of win and loss. The main lesson is that where truly paradoxical consequences of SIA are concerned one cannot legitimately "pass-the-buck" to the general probability/game theory. The conclusion that SIA apparently supports all sorts of lunatic theories as long as they imply large number of observers still stands. Although it is still to some degree a matter of taste and (probabilistic) guts, such price for escaping the controversial consequences of DA seems too high to the present author.

The claim made by Olum (2001) that the eternal inflation and similar chaotic multiverse theories successfully obviate DA is also highly suspicious. It is based on at least two controversial premises:

1. The no-outsider requirement (Bostrom 2000, p. 123), which is by no means indisputable, and in the Leslie's (1996) version of DA is absent.
2. The prejudice according to which observers in different universes must (or even may) belong to the relevant reference class.

Concerning the no-outsider requirement, it seems weakly supported and should be eliminated alltogether, but the detailed discussion is beyond the scope of the present manuscript. The opinion of the present author is that in the light of modern cosmology, the no-outsider requirement undermines **any** form of anthropic reasoning, since we expect that there will **always** be an actual infinite number of observers in the universe (or multiverse, in the cosmological sense); rejection of this requirement is hence desirable, even if it comes with a price of a modest form of anthropocentrism (manuscript in preparation). The second assumption is controversial from several points of view. If one of the basic motivations for believing in the (cosmological) multiverse structure of reality is empirically established fine-tuning of our universe, this immediately implies that other universes are mainly uninhabitable. If we accept the estimate of Penrose (1989), one expects a universe similar to ours once in every $10^{10^{123}}$ uninhabitable universes. Now the presumptive character of SIA is manifested on even more grandiose scale than in the case of our old friend, the presumptuous philosopher. It would be bad form of anthropocentrism to conclude that standards of observership valid here are necessarily valid in the other universes with different physical laws and different sets of fundamental constants—especially if this far-fetched conclusion is motivated by desire to avoid what is perceived as the controversial conclusion of DA. Quite to the contrary, there may be some reason following from the foundations of the decoherence theory to expect that different universes will determine different forms of consciousness, and therefore entail radically different standards of observership (Dugić et al. 2000). Such approach, exactly contrary to the one intended to refute DA in the multiverse case, will have an additional advantage in automatically solving the reference class problems, at least for this case.

In addition, one may recognize a double standard in trying to solve the DA "problem" in the multiverse, since if Bostrom (2000) is incorrect in applying SSA to causally disconnected regions **of our universe**, the same verdict applies *a fortiori* to any pronouncement taking into account inhabitants of **other universes**. In particular it is so as long as no satisfactory theoretical argument or observational finding for a topological connection between various universes is offered. The least one can say as a conclusion to this issue is that postulating the existence (and particular properties) of the multiverse can not really help us solve the DA problem.

## 6. Conclusions: Too Early for Optimism?

Our conclusion is that it is too early to accept SIA as an antidote to controversial and (seemingly) apocalyptic conclusions of the Doomsday argument. We know too little on the necessary physical requirements for observership to establish what additional assumptions are more justified in application of statistics to intelligent observership on that (physical, chemical, etc.) grounds. While pursuing that goal, we should also seriously consider epistemological and methodological grounds for accepting or rejecting any of these general assumptions proposed in the literature (SSA, SIA, SSSA, etc.). In particular, the Self-Indication Assumption is unable to satisfactorily substitute for our ignorance in the relevant field. It has several empirically wrong and epistemologically contrived consequences, while there is no compelling evidence for accepting it—apart from the DA cancellation. On the other hand, it runs afoul of "no-collapse" quantum mechanical theories, as well as several cosmological and astrobiological arguments (notably Davies-Tipler argument). The Presumptuous Philosopher thought experiment remains a serious problem for it, and may be cast in even more detrimental form, from the SIA proponents' point of view. Finally, issues in confirmation theory become hopelessly muddled whenever—SIA accepted—theories involving large number of observers are invoked. Taken together, this argumentation indicates the insufficiency of Self-Indication Assumption and the necessity of taking some different path if the refutation of Doomsday Argument is desired.

**Acknowledgements.** Among people with whom author have discussed the topics relevant to this manuscript names of Ken Olum, Fred Adams, Miloš Arsenijević, Maja Bulatović, Petar Grujić should be mentioned. Special thanks are due to Nick Bostrom, who offered not only his philosophical insights, but also a kind and wholehearted support.

[1] For instance, p. 316 of Gott (1993).

[2] These examples are elaborated in Bostrom (2002). The present author would add importance of these assumptions for the search for extraterrestrial intelligence (SETI) problems and—rather obviously—demographics.

[3] That B-theories of time ("tenseless", "atemporal") are more in accordance with the findings of physical sciences has been commented by Grünbaum (1991), Oaklender (1984) and Price (1996). Also, the general plausibility of backward causation has been recently defended, among others, by Brown (1992).

[4] On the contrary, one may argue that if there is a causal problem, it may obviate the **reasoning** linking the sampling assumption to "paradoxical" consequences, and not the assumption itself. For instance, we might claim that the very validity of SSA implies "doomsday" of a sort which will prevent the creation of a governing body such as the described UN$^{++}$ (a manuscript of the present author, in preparation).

[5] Thus, for instance, Prof. William L. Craig writes: "It is entirely possible that the time of physics *is*, in fact, a B-theoretic time, but that this is an abstraction, a skeleton, of full-blooded metaphysical time, which is an A-theoretic time." (Craig 1992)

[6] A nice semi-popular exposition of the cosmological horizon lore can be found in Ellis and Rothman (1993).

[7] As is well known, $\Omega$ is the ratio of physical density of all matter fields to the so-called critical density necessary for universe to stop current expansion and recollapse toward Big Crunch. Thus, $\Omega \leq 1$ universes will expand forever, while $\Omega > 1$ universes will recollapse. This strictly applies only to the case of matter fields possessing "regular" equation of state; in the presence of vacuum energy, indicated by recent cosmological supernovae experiments, the situation becomes more complicated. However, the basic conclusion that if $\Omega$ in the matter fields ($\Omega_m$) is less than (or equal to) unity and the sign of additional vacuum energy is positive (which is the standard view) the universe is infinite and ever-expanding.

[8] The proponent of SIA could argue against such conclusion by stating that—especially in view of the theory-like nature of the "no-collapse" view (see below and the next endnote)—the condition of *ceteris paribus* is never satisfied. However, this argumentation brings us in an awkward epistemic position since then one may also argue that yet unobserved minuscule effects (like those of inter-branch interaction which may manifest themselves in the Everett's theory and not in the standard quantum mechanics) will **always** obstruct using SIA or any other probabilistic assumption in evaluation of scientific theories. This recourse amounts to a sort of global explanatory nihilism to be accepted only in the complete lack of alternatives.

[9] And there is a slowly building consensus that it is a separate theory, not just an interpretation of the quantum mechanical formalism; for some of the arguments, see Squires (1986), Price (1996) and Page (1999). Some of the (still thought) experiments discriminating between Everett's theory and the orthodox quantum mechanics are described by Deutsch (1985), Plaga (1997) and Tegmark (1998). This also comes some further steps toward addressing Olum's correct point that "physical processes (such as those that might lead to destruction) should not depend on a choice of quantum mechanics interpretations"; however, it is hardly possible to claim that **predicted** physical processes should not depend on a choice of the correct quantum mechanical **theory**!

[10] There is an additional tacit assumption here, and that is that the **topology** of our universe is either trivial, or at least not contrived in such way that arbitrarily high negentropy sources in other regions ("other universes") may be brought into contact with the heat bath in a fixed comoving volume. This assumption is a rather weak one, and even if it is not entirely satisfied, one may argue that there are extraneous requirements that limit the growth of civilizations in such a universe (Ćirković and Bostrom 2000).

[11] This is the famous argument of Feynman (1965) against the anthropic explanation of the thermodynamical temporal asymmetry. As Barrow and Tipler (1986) correctly point out, it is not entirely valid in the context of modern cosmology, and there is some minimal size which **any** universe containing intelligent observers must have, due to power spectrum of density perturbations, galaxy formation, and other astrophysical "details". However, these latter considerations should not worry SIA-proponents: they will find Feynman's remark *a priori* wrong, since it predicts only a very small number of observers in the universe both at present and at all times (while there is at least one competing empirically acceptable theory implying a larger number of observers). If only solving cosmological puzzles were that easy!

[12] Contrary to the situation encountered in most philosophical discourses, even recent ones (e.g. Price 1996), practically entire observational cosmology offers arguments supporting the infinite size of the universe (e.g. Peebles 1993). The conclusion that our universe is open (i.e. that the density of the matter in the universe is significantly smaller than the critical value for recollapse, or at best equal to it) is strongly supported by a recent tentative discovery of large positive cosmological constant (Perlmutter et al. 1999; Krauss and Turner 1999). Another important consequence of this discovery is the existence of the cosmological **event** horizon, as different from usually considered **particle** horizon (as discussed above).